\documentclass{cernrep} 
\usepackage{texnames}
\usepackage[T1]{fontenc}
\begin{document}
\title{Evolutionary Computation in High Energy Physics}
 
\author{Liliana Teodorescu}

\institute{Brunel University, United Kingdom}

\maketitle
 
\begin{abstract}
Evolutionary Computation is a branch of computer science with which, traditionally,
High Energy Physics has fewer connections. Its methods were investigated
in this field, mainly for data analysis tasks. These
methods  and studies are, however, less known  in the high energy physics community 
and this motivated us to prepare this lecture.  The
lecture presents a general overview of the main types of algorithms based on
Evolutionary Computation, as well as a review of their applications in High Energy Physics. 

\end{abstract}
 
\section{Introduction}

\hspace{0.9cm}Evolutionary Computation is a branch of computer science which  aims to 
develop efficient computer algorithms for solving complex problems by modelling the natural 
evolution.

Natural evolution, in this context, is defined as the 
optimisation process which aims to increase the ability of individuals to survive and 
reproduce in a specific environment. This ability is quantitatively measured by
the evolutionary fitness of the individuals. The unique characteristics of each individual 
are represented in its chromosomes. Through the natural selection process the fitter
 chromosomes will mate more often creating offspring with similar or better fitness.
The goal of the natural evolution process is to create a population 
of increasing fitness.

The algorithms based on Evolutionary Computation, called Evolutionary Algorithms,
 use simulation of natural evolution on a computer. The candidate solution of the problem 
to be solved by the algorithm represents an individual which is encoded in a form understood
 by the computer and called a chromosome. A chromosome can be divided into one or more 
constituent parts called genes. The quality of the candidate solution is evaluated with an
objective function called the fitness function. The reproduction process of the individuals is simulated 
by applying on them a set of operators, called genetic operators,  creating genetic variation.
The selection of the individuals for reproduction is, usually, proportional to their 
fitness (the value of the fitness function for the individual). Through an iterative process, 
the algorithm improves the quality of the solution until an optimal solution is found.

Evolutionary Algorithms (EA) refer, traditionally, to four main types of algorithms:
Genetic Algorithms (GA) \cite{holl75}, Genetic Programming (GP) \cite{koza92}, 
Evolutionary Strategies (ES) \cite{schw81} and Evolutionary Programming (EP) \cite{foge66}.
More recent developments, such as Gene Expression Programming (GEP) \cite{ferr13},
combine some characteristics of the previous versions. 

In high energy physics, EA were successfully tested but not largely used,
mainly due to their high computational needs. GA were applied mainly to problems
such as discrimination and parameter optimisation in both experimental and theoretical studies 
(for example, see \cite{ire04,all04,abd03,azu98}). GP 
was recently applied to event selection type problems \cite{atlas,foc0503,foc0507}. 
ES were tested for optimisation of event selection criteria \cite{berl04}.
The first application of GEP to a high energy physics problem was presented in
\cite{ieee05}. Comparative studies of event selection with GEP and with commonly used methods 
in high energy physics were presented in \cite{acat07}.

This lecture introduces the basics of EA, presents the main classes of such algorithms
and summarises their applications in high energy physics. Mode detailed presentations of EA
can be found in textbooks such as \cite{book-ci}, \cite{banz}, \cite{ec-book}, 
\cite{ga-book}.

\section{Evolutionary Algorithms}

\subsection{Basic structure of an evolutionary algorithm}

\hspace{1.0cm}The first steps, and the most difficult ones, of the application  of
any EA are the problem definition, the encoding of the candidate solution and the
definition of the fitness function. The encoding and the fitness functions are 
specific to each problem. Adequate choices are crucial for the success of the
algorithm. In making these choices  knowledge about the problem and about the
expected solution should be used. 

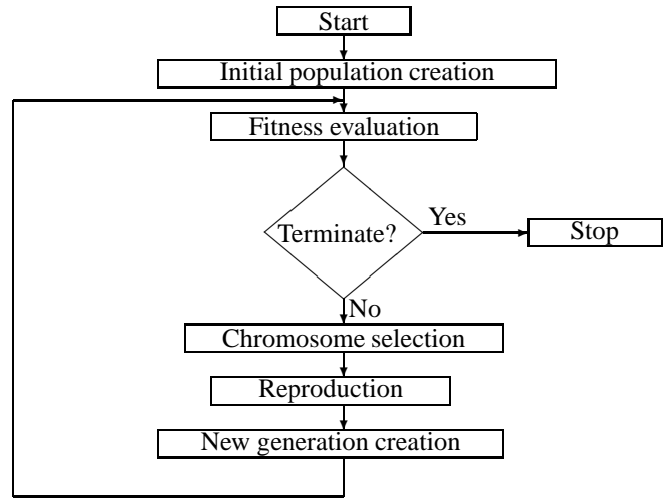
\begin{figure}
\centering
\begin{picture}(180,185)
\fontsize{10}{12}\selectfont
\put(65,175){\framebox(50,10){Start}}
\put(20,155){\framebox(150,10){Initial population creation}}
\put(40,135){\framebox(100,10){Fitness evaluation}}
\put(160,95){\framebox(50,10){Stop}}
\put(30,55){\framebox(120,10){Chromosome selection}}
\put(40,35){\framebox(90,10){Reproduction}}
\put(20,15){\framebox(130,10){New generation creation}}
\put(90,15){\line(0,-1){15}}
\put(90,0){\line(-1,0){125}}
\put(-35,0){\line(0,1){150}}
\put(-35,150){\vector(1,0){125}}
\put(90,125){\line(6,-5){30}}
\put(90,125){\line(-6,-5){30}}
\put(90,75){\line(6,5){30}}
\put(90,75){\line(-6,5){30}}
\put(65,96.5){Terminate?}
\put(90,175){\vector(0,-1){10}}
\put(90,155){\vector(0,-1){10}}
\put(90,135){\vector(0,-1){10}}
\put(90,75){\vector(0,-1){10}}
\put(90,55){\vector(0,-1){10}}
\put(90,35){\vector(0,-1){10}}
\put(120,100){\vector(1,0){40}}
\put(92,68){No}
\put(122,103){Yes}
\end{picture}
\caption{Basic EA algorithm}
\label{algorithm}
\end{figure}

The next step is the application of the algorithm itself. A basic representation of an EA
 is presented in Fig. \ref{algorithm}.

The actual running of the algorithm starts with the random creation of an initial population
of chromosomes. The fitness function is then
evaluated for each chromosome determining the chromosome's fitness. Using this fitness, 
a termination criterion is evaluated indicating if a solution of the desired quality 
was found or a certain number of iterations was run. If the termination
criterion is not met, some of the chromosomes are selected and reproduced, resulting in offspring.
The new chromosomes will replace the old ones producing a new generation. The process
continues until the termination criterion is met. The fittest chromosome is then
decoded, producing the optimal solution of the problem as it was developed by the algorithm.

\subsection{Solution representation}

\hspace{1.0cm} A candidate solution of the problem at hand is represented
by a chromosome. In this respect, a chromosome represents a point in the search space.

Choosing an adequate chromosome representation is very important for the success of the algorithm
as it will influence the efficiency and the complexity of the searching process.

Different types of EA use different solution representation schemes. 
Commonly used are schemes based on fixed or variable-length strings (GA, EA and EP), or
 on trees (GP). Binary strings with each bit representing  a boolean value, an integer or a 
discretised real number (GA), or  strings of real-valued variables (ES, EP) can be used.
 A combination of the string and tree representations  is used  by GEP. In this case the chromosome 
containing the genetic material
(information from which the candidate solution is built) is represented by a string and
mapped into a tree which represents the actual candidate solution.

\subsection{Fitness function}

\hspace{1cm} The fitness function maps the chromosome representation into a scalar value.
It describes how good a candidate solution is for the problem at hand. For this reason, it is
one of the most important component of an evolutionary algorithm.

The fitness function has to contain all the criteria which need to be optimised in the searching process,
and to reflect the constrains of the problem. Its value (fitness value) is used in the selection
of the chromosome for reproduction, as well as in defining the probabilities with which
the genetic operators are applied.

\subsection{Initial population}

\hspace{1cm} The most common way of generating the initial population is by generating
random values for each gene of the chromosomes from the allowed set of values. In this way,
a uniform representation of the search space is ensured.

If {\it a priori} knowledge about the solution exists, the initial population can be biased 
towards potentially good solutions. This involves, however, a certain risk for a premature 
convergence to a local optimum. 

The size of the initial population is chosen by the user. The size of the population will
influence the time complexity per generation and the number of generations needed to reach
 the convergence of the algorithm. Small generations imply low time complexity per generation 
but  more generations are needed for reaching convergence.

\subsection{Selection operators}

\hspace{1cm} Selection operators are used to select individuals for applying genetic operators
and for creating the new generation. Various selection methods exists. There is no clear evidence
of any method being superior. 

A few common selection methods are:
\begin{itemize}
\item{random selection} - individuals are selected randomly, without any reference to fitness;
\item{proportional selection} - the probability to select an individual is proportional to the
fitness value; when the fitness value is normalised to the maximum fitness, the method is known
as the roulette wheel method;
\item{rank-based selection} - the probability to select an individual is proportional to the
rank order of the fitness value, rather than to the fitness value itself.
\end{itemize}

\subsection{Reproduction operators}

\hspace{1cm}Reproduction or genetic operators are applied on the selected individuals in order to create
offspring which will constitute the next generation. Typical genetic operators are:

\begin{itemize}
\item{cross-over} - combines genetic material of two parents, producing two new individuals;
\item{mutation} - randomly changes the values of genes in the chromosome, introducing new genetic
material; in order not to disturb the good genetic material, this operator has to be applied
with low probability;
\item{elitism or cloning} - copies the best individuals in the next generation, without any modification.
\end{itemize}

The exact structure of the genetic operators is specific to each chromosome representation, and hence to each
type of EA. Examples will be given in the next sections.

\section{Genetic Algorithms}

\subsection{Algorithm}

\hspace{1cm} Genetic Algorithms were proposed by J.H. Holland in 1975 \cite{holl75}.
The  classical version of the algorithm uses a chromosome representation based on a
binary string of fixed length. A chromosome consists of a set of variables encoded 
as a binary string. These variables could be binary variables (each variable encoded
as a bit),  nominal-valued discrete variables (each nominal value encoded
as a bit string) or countinous-valued variables (each variable mapped to a bit string
with a certain algorithm).Alternative chromosome representations were subsequently proposed
such as integer or real-valued representations, order-based representations, chromosomes
of variables length and many more \cite{hand}.

\begin{figure}[!htbp]
\begin{center}
\includegraphics[width=10cm,height=5cm]{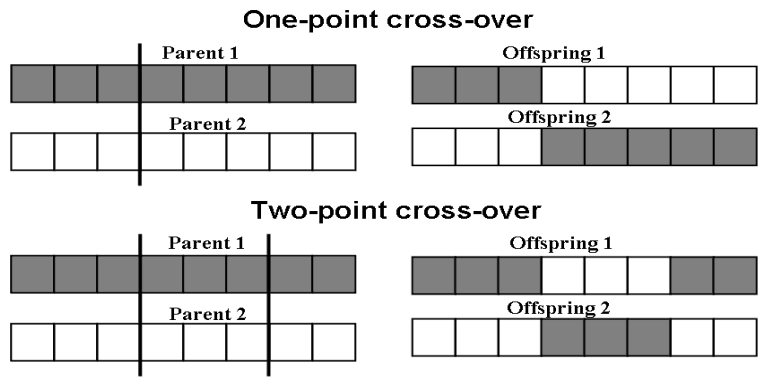}
\end{center}
\caption{One-point and two-point cross-over operators in GA}
\label{cross-over-ga}
\end{figure}

Reproduction is traditionally made  with the cross-over and mutation operators.

The cross-over operator is applied on a pair of chromosomes by exchanging genetic
 material between them and creating two new chromosomes. The operator is applied
with a certain probability called the cross-over rate.
Two common versions of the operator are  one-point and two-point cross-over.
One-point cross-over exchanges the genetic material between
one randomly chosen point in the chromosomes and their ends.
Two-point cross-over exchanges the genetic material between two points randomly
chosen in the chromosomes. An example of how the two operators work
is shown in Fig. \ref{cross-over-ga}.

\begin{figure}[!htbp]
\begin{center}
\includegraphics[width=5cm]{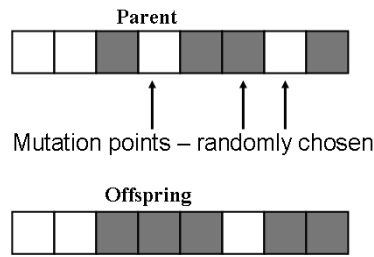}
\end{center}
\caption{Mutation operator in GA}
\label{mutation-ga}
\end{figure}

Mutation operator changes the value of randomly chosen genes in the chromosome.
It is applied with a small probability, called mutation rate, in order to 
avoid a massive destruction of the good genetic material in the population. 
An example of how the mutation operator works in GA is shown in Fig. \ref{mutation-ga}.   

\subsection{Applications in high energy physics}

\hspace{1cm}In high energy physics, GA were investigated for large-scale parameter optimisation
and fitting problems in both experimental and phenomenological physics.

In experimental high energy physics problems like cut value optimisation for
event selection \cite{ga-acat07}, trigger optimisation \cite{abd03}
or neural-network parameter optimisation for event selection  \cite{ga-nn} 
with GA were studied. 

In \cite{ga-acat07}, for example, a software implementation of GA adapted for the optimisation
of the cut values of event variables was presented. Tests on simulated data showed
improved optimised cut values produced by the algorithm. The effective gain, comparative
to a manual optimisation, depends on the chracteristics of the data and of
the physics analysis performed.  

In phenomenological high energy physics GA were used for optimising various parameters
of the theoretical models as, for example, in isobar models for $p(\gamma K^*)\Lambda$
reaction \cite{ire04}, SUSY model discrimination \cite{all04} or lattice calculations
\cite{azu98}.

In \cite{all04}, for example, GA was used to address the problem of discriminating SUSY
breaking models by evaluating the accuracy on measurements required to distinguish two
models. Three SUSY breaking scenarious were used in this study considering mass as
the relevant observable. The fitness function was defined as a ``relative distance''
which describe the relative difference between two mass spectra. The procedure 
indicated values of the ``relative distance'' of the order of 1\% which means the
rough accuracy required for the sparticle mass measurements should be of this
order for allowing the discrimination of the theoretical models.

\section{Evolutionary strategies}

\subsection{Algorithm}

\hspace{1cm} Evolutionary Strategies were proposed by I. Rechenberg in 1973 \cite{rech73}
and developed further by H-P. Schwefel in 1975 \cite{sch75}. They use the concept of evolution 
of evolution. Evolution is considered a process which optimises itself as the result of the 
interaction with the environment.

In order to implement this concept, an individual is represented by its  genetic 
material and a so called strategy parameter which models the behavior 
of the individual in the environment:

\begin{equation}
C_{g,i} = (G_{g,i},\sigma_{g,i}),
\end{equation}

where $C_{g,i}$ is the individual $i$ of the generation $g$,
 $G_{g,i}$ is the genetic material of the individual and $\sigma_{g,i}$ is its strategy parameter.  

The genetic material is represented by floating-point variables. The strategy parameter 
is, usually, the standard deviation of a normal distribution associated with each individual or
with each variable of the individual.

The evolution process  evolves both the genetic material  and the strategy parameter. 

The initial version of ES used only mutation in the reproduction process. Other versions
were subsequently developed using both cross-over and mutation ( for a short description
and more references, see \cite{book-ci}). 

The cross-over operator is implemented as  a local operator (randomly selected material from 
two parents is selected and recombined to form a new chromosome) or as a global operator (randomly
selected material from the entire generation is selected and recombined to form a new
chromosome). Two types of recombinations are commonly used:
 discrete recombination in which the gene value of the offspring is the gene value from the 
parent, and  intermediate recombination in which  the midpoint between the gene value of the 
parents gives the gene value of the offspring.

Mutation operator has a special implementation in ES. It will mutate both the strategy parameter
$\sigma_{g,i}$ and the genetic material $G_{g,i}$ by modifying them with the following
relations:

\begin{equation}
\sigma_{g+1,i}=\sigma_{g,i}e^{\tau \zeta_{\tau}},
\end{equation}

\begin{equation}
G_{g+1,i}=G_{g,i}+\sigma_{g+1,i}\zeta,
\end{equation}

where $\tau = \sqrt{I}$, $I$ being the chromosome dimension (number of genetic variables),
$\zeta_{\tau}$ and $\zeta \sim N(0,1)$.

In ES and  in other EA the
offspring resulting from mutation is accepted only if it is fitter.

\subsection{Applications in high energy physics}

\hspace{1cm} To the best knowledge of the author of this lecture, there is only one
study of the  ES applicability to high energy physics problems \cite{berl04} in which
 the algorithm was tested for optimisation of event selection cuts and for 
parameter minimisation in a Dalitz plot analysis. 

Using the squared signal significance ($S^2/(S+2B)$, $S$ - number of signal events, 
$B$- number of background events) as a fitness function and four input variables 
(momentum of the reconstructed
system - $D_s$ in this study, the width of the mass window around the reconstructed 
system, the probability of the vertex fit and the helicity angle) an ES algorithm was used
to optimise the values of the cuts applied on these four variables such that
 the signal significance is maximised. The procedure was applied for the event selection 
corresponding to the
following decay processes studied in the BaBar experiments \cite{babar}: $D_s \rightarrow \phi \pi$,
$D_s \rightarrow \overline{K}^{*0}K^+$  and $D_s \rightarrow \overline{K}^0K$.
Starting with values optimised manually, the algorithm found cut values which improved
the squared signal significance for the three processes by 19.4\%, 45\% and 16\%, respectively.

For the Dalitz plot analysis, ES was used to optimise the weights of the MC events such that 
the difference between the MC and data events in the Dalitz plot for the reaction
$p\gamma \rightarrow \pi^0 \eta$ (studied by the CB/ELSA collaboration \cite{elsa}) was minimised. 
In this study 16 parameters were optimised. The resulting MC Dalitz plot  
was in good agreement with the experimental data. 

\section{Genetic Programming}

\subsection{Algorithm}

\begin{figure}[!htbp]
\begin{center}
\includegraphics[width=6cm,height=2cm]{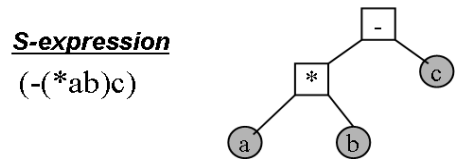}
\end{center}
\caption{An S-expression and its corresponding GP tree }
\label{tree-gp}
\end{figure}

Genetic programming was proposed by J.R. Koza in 1992 \cite{koza92}.
In contrast with  the previously discussed EA, GP searches for the computer program
which solves the problem at hand rather than for the solution to the problem.
While the initial intentions and hopes were GP to be developed for  generating 
computer programs in any computer language, it was only used for generating
computer programs as S-expressions in LISP which are graphically translated as trees, 
as shown in Fig. \ref{tree-gp}.
In this figure a,b,c are variables or constants and are called terminals. They
are used together with mathematical functions in forming the S-expressions or the GP trees.

In GP the chromosome is represented as a tree of variable length. The variable
length gives more flexibility to the algorithm. There are, instead, syntax constraints
which need to be satisfied. The unprotected evolution generates many invalid
trees which need to be eliminated, resulting in waste of CPU resources.

Reproduction is performed in GP mainly with the cross-over and mutation operators. 
They have specific implementations, adapted to the tree representation.

\begin{figure}[!htbp]
\begin{center}
\includegraphics[width=12cm,height=8cm]{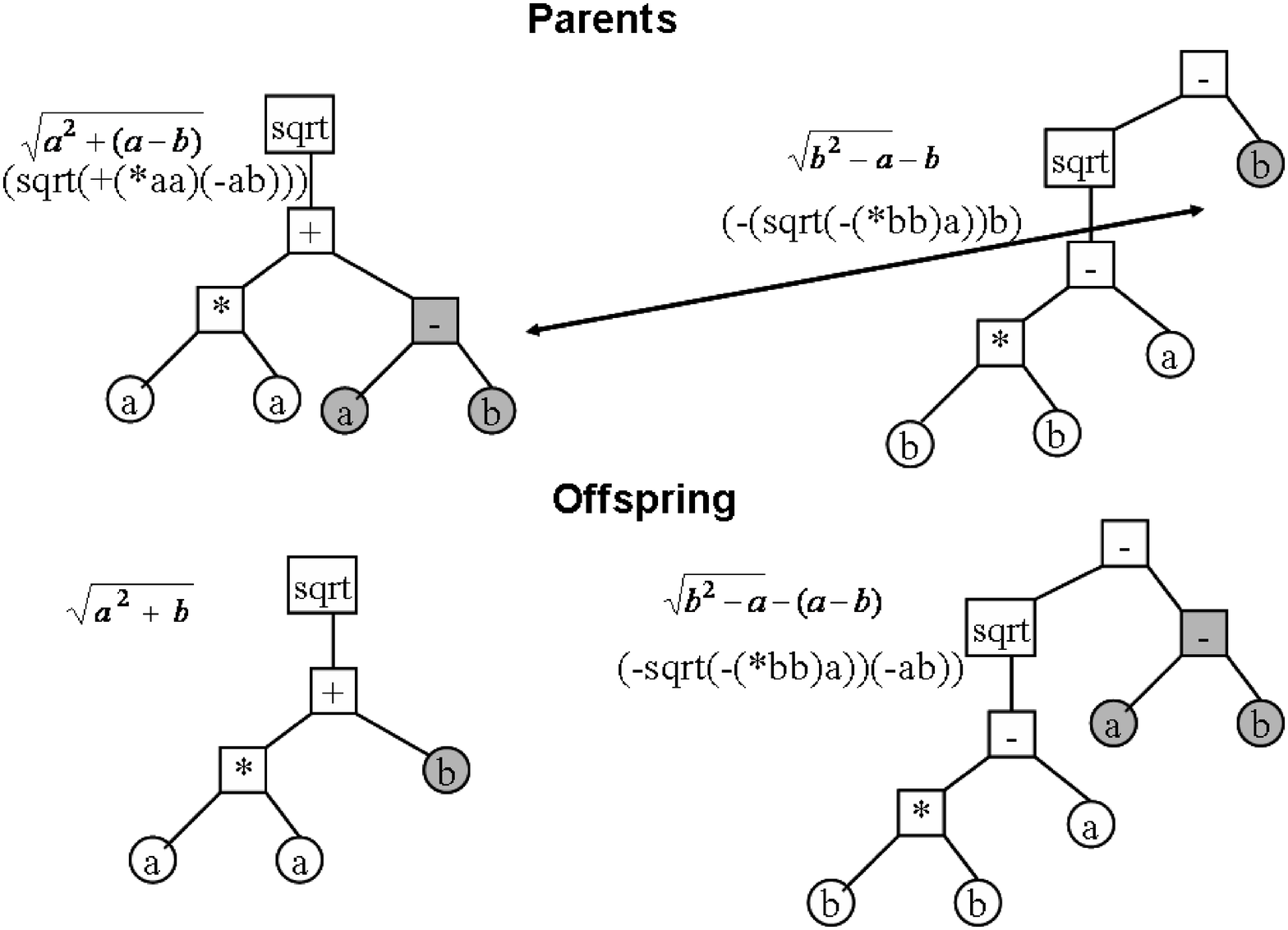}
\end{center}
\caption{Cross-over operator in GP}
\label{cross-over-gp}
\end{figure}

The cross-over operator exchanges parts of two parent trees resulting in two new
trees. An example is shown in Fig. \ref{cross-over-gp} which displays the 
parent and the offspring trees, as well as the corresponding S-expressions and
mathematical expressions.

\begin{figure}[!htbp]
\begin{center}
\includegraphics[width=12cm,height=8cm]{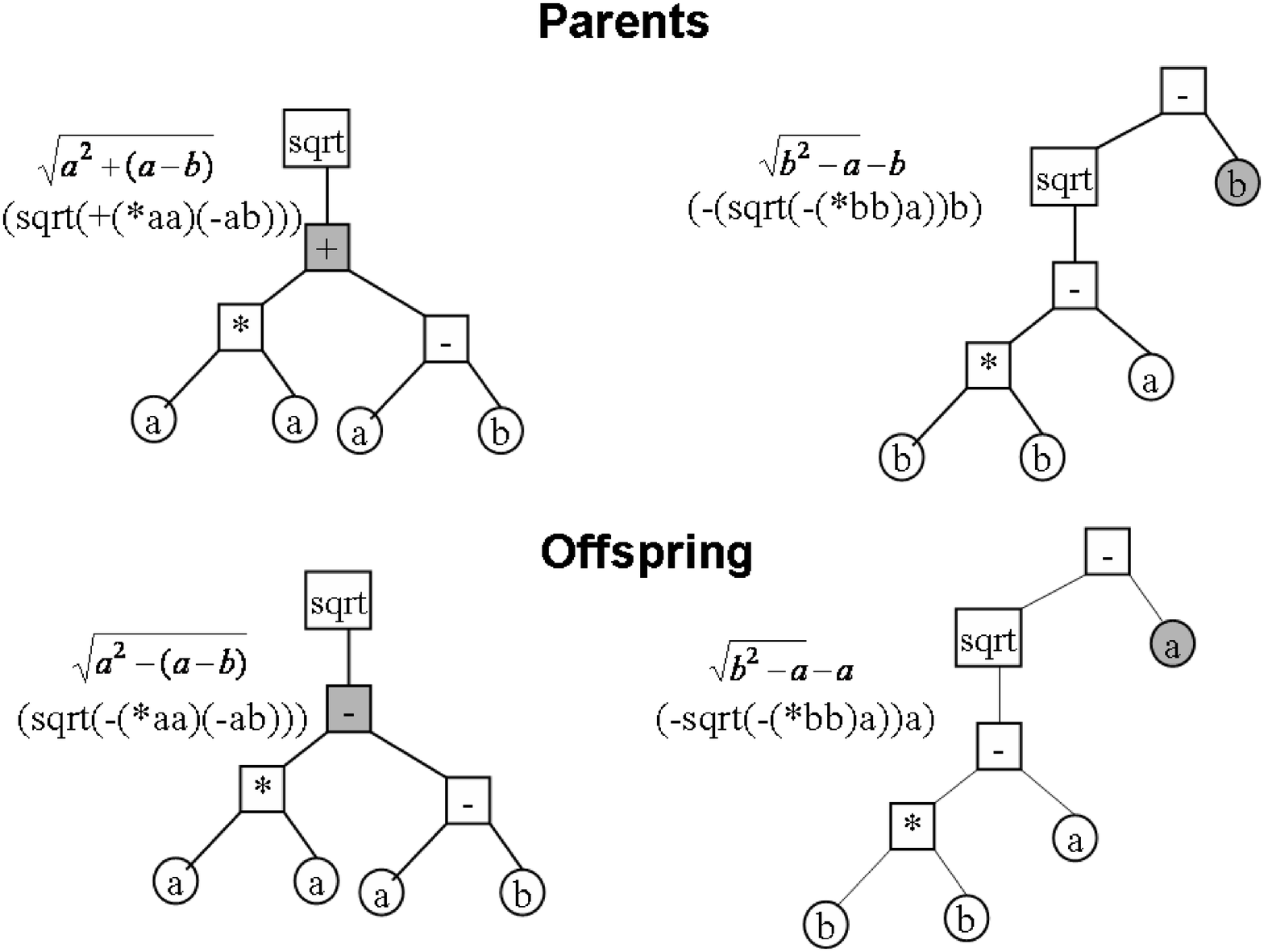}
\end{center}
\caption{Mutation operator in GP}
\label{mutation-gp}
\end{figure}

The mutation operator changes a function of the tree into another function, or
a terminal into another terminal. An example is shown in Fig. \ref{mutation-gp}
 which displays the 
parent and the offspring trees, with the corresponding S-expressions and
mathematical expressions, for the two types of mutation.

In order to deal with the syntax constraints, various versions of GP were developed
by imposing certain grammar checks in the evolution process.

\subsection{Applications in high energy physics}

\hspace{1cm} GP was only recently tested for data analysis tasks in high energy
physics. The FOCUS experiment \cite{focus} developed a methodology for event selection with GP
and applied it to the analysis of its experimental data for the following
processes: $D^+ \rightarrow K^+\pi^+\pi^-$, $\Lambda_c \rightarrow pK^+\pi^-$
and $D_s^+ \rightarrow K^+K^+\pi^-$ \cite{foc0503}, \cite{foc0507}.
A similar study for Higgs searches in ATLAS experiment was presented in \cite{atlas}
using Monte Carlo simulated data.

As an example, the procedure developed by the FOCUS collaboration is shortly summarised here.

The chromosome is built from candidate cuts for separating the signal events from 
background events for the desired 
physics process. It is represented by a tree built from a set of functions (common
mathematical functions and operators, and  boolean operators), a set of event variables
(vertexing, kinematics and particle identification variables) and constants created
by the algorithm in the range \mbox{(-2.0,+2.0)} for real constants and (-10,+10) for integer
constants.

The following fitness function was  minimised by the algorithm:

\begin{equation}
\frac{S+B}{S^2} \times 10000 \times (1+0.05\times n),
\end{equation}

where $S$ and $B$ represents the number of signal and background events, respectively,
and $n$ is the number of nodes in the GP tree.

The term proportional to $n$ was introduced in order to control the number
of nodes in the tree such that the big trees would have a high chance to survive in the
evolution process only if they make a significant contribution to the background
reduction or to the signal increase. 

The basic procedure of the analysis was:
\begin{itemize}
\item an initial generation of chromosomes was  almost randomly created,
\item for all events in the sample the fitness value of each chromosome was calculated.
Only the events for which the tree is evaluated to give a positive value were kept.
For the surviving events, the  invariant mass distribution of the studied system was
fit with adequate functions and the number of signal and background events was determined. 
\item the chromosomes were selected, modified with genetic operators and a new 
generation created,
\item the process was repeated for the desired number of generations.
\end{itemize} 

The best chromosome developed at the end of the process represented the final selection
criteria to be applied on the data for selecting the desired physics events. It had the 
shape of a quite complex tree, with 38 nodes. 

The procedure was validated by comparing the results, in terms of events yield, with
those obtained with a standard analysis, based on rectangular cuts applied on the
event variables. Similar results were obtained in the two analyses.

\section{Gene Expression Programming}

\subsection{Algorithm}

\hspace{1cm} Gene Expression Programming was proposed by C. Ferreira in 2001 \cite{ferr13}.
The main difference between GEP and the other EA discussed here is the separation of the 
solution representation into two parts: the chromosome which encodes the information
used in building the solution, and the expression tree (ET) which represents the
candidate solution itself. Such separation, using different representations, was
proposed before as, for example, in  Developmental Genetic Programming (DGP) \cite{banz}
with which GEP has important similarities.

 The GEP chromosome is a list of functions and terminals (variables and constants) organised
in one or more genes of equal length. The functions and variables are input information
while the constants are created by the algorithm in a range chosen by the user.
 Each gene is divided into a head composed of terminals and functions, and a tail composed 
only of terminals. The length of the head ($h$) is an input parameter of the algorithm while 
the length of the tail ($t$) is given by:

\begin{equation}
t=h(n-1)+1,
\label{len}
\end{equation}

where $n$ is the largest arity of the functions used in the gene's head.

This head-tail partition of the gene ensures that every function of the gene has the
required number of arguments available, making the chromosome correspond to a syntactically 
correct expression.

Each gene of a chromosome is translated (decoded) into an ET with the following rules:

\begin{itemize}
\item the first element of the gene is placed on the first line of the ET and constitutes
its root,
\item on each next line of the ET a number of elements equal to the number of arguments of the 
functions located on the previous line is placed,
\item the process is repeated until a line containing only terminals is formed. 
\end{itemize}

The reverse process, the encoding of the ET into a gene, implies reading the ET from
left to right and from top to bottom.

\begin{figure}
\centering
{
\begin{picture}(180,200)(0,-30)
\put(10,160){\it Chromosome with one gene}
\put(15,145){{\bf *b+a-aQab+//+b+}babbabbbababbaaa}
\put(45,133){head} \put(150,133){tail}
\put(15,130){\line(1,0){225}} \put(135,130){\vector(0,1){14}}
\put(15,105){\it Expression tree}
\put(20,70){\circle{14}} \put(25,75){\line(1,1){10}} \put(17.4,67){\bf b}
\put(40,90){\circle{14}} \put(45,85){\line(1,-1){10}} \put(37.4,85.5){\bf *}
\put(40,50){\circle{14}} \put(45,55){\line(1,1){10}} \put(37.4,47.5){\bf a}
\put(60,70){\circle{14}} \put(65,65){\line(1,-1){10}} \put(57,67.5){\bf +}
\put(60,30){\circle{14}} \put(65,35){\line(1,1){10}} \put(57.4,27.5){\bf a}
\put(80,50){\circle{14}} \put(85,45){\line(1,-1){10}} \put(74,46.3){ \LARGE -}
\put(100,30){\circle{14}} \put(100,23){\line(0,-1){6}} \put(96,27){\bf Q}
\put(100,10){\circle{14}} \put(97.4,7.5){\bf a}
\put(15,-15){\it Mathematical expression}
\put(20,-30){$b*(a+(a-\sqrt{a}))$ = $b*(2a-\sqrt{a})$}
\end{picture}}
\caption{Unigenic chromosome, the decoded ET and its corresponding mathematical expression}
\label{gene}
\end{figure}

An example of a chromosome with the head length equal to 15  made of
five functions, $Q,*,/,+$ and $-$, ($Q$ being the square root function) and two terminals,
a and b, is shown in 
Fig. \ref{gene},
 together with its decoded ET and the corresponding mathematical expression.

In the case of multigenic chromosomes, the ETs corresponding to each gene
are connected with  a linking function defined by the user. The mathematical expression
associated with these combined ETs is the candidate solution to the problem.

The reproduction process takes place by applying genetic operators on the chromosome
(not on ET, as in GP). GEP uses four types of genetic operators:

\begin{figure}[!htbp]
\begin{center}
\includegraphics[width=12cm,height=5cm]{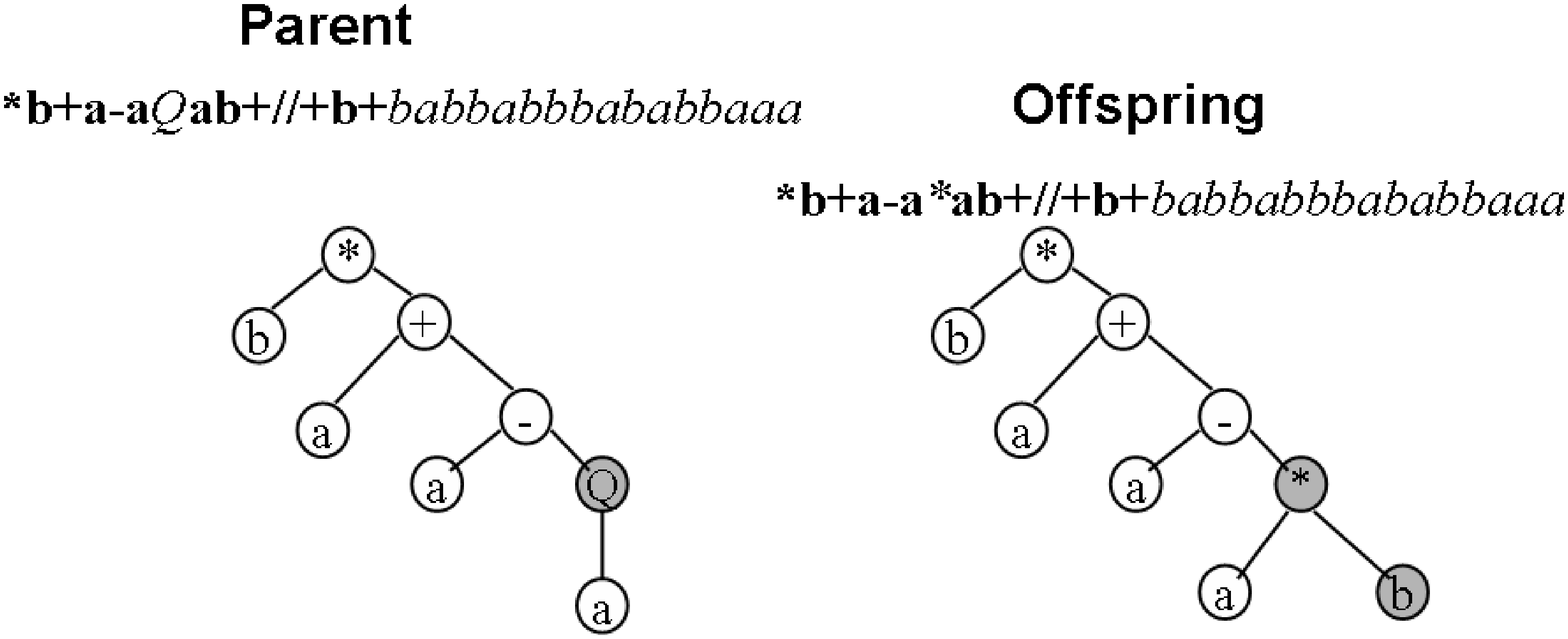}
\end{center}
\caption{Mutation operator in GEP}
\label{mutation-gep}
\end{figure}

\begin{itemize}
\item elitism - the fittest chromosome is replicated unchanged into the next generation, 
preserving the best material from one generation to another.
\item cross-over - exchanges parts of a pair of randomly chosen chromosomes.
In addition of the one-point and two-point cross-over, as in GA, GEP uses gene
cross-over in which entire genes are exchanged between two parent chromosomes, 
forming two new  chromosomes containing genes from both parents.
\item  mutation - randomly changes an element of a chromosome into another element,
preserving the rule that the tails contain only terminals.
In the head of the gene a function can be changed into another function or terminal and 
{\it vice versa}. In the tail a terminal can only be changed into another terminal.
\item transposition - randomly moves a part of the chromosome to another location
in the same chromosome.
\end{itemize}

An example of how such an operator (mutation, for example) works in GEP is shown in 
Fig. \ref{mutation-gep}.

\subsection{Applications in high energy physics}

\hspace{1cm} The author of this lecture performed the first study of the applicability of 
GEP to an event selection problem in high energy physics. The selection of
$K_S$ particle produced in $e^+e^-$ interaction at $10GeV$ and reconstructed in the
decay mode $K_S \rightarrow \pi^+ \pi^-$  was used as an example application 
\cite{ieee05}, \cite{acat07}. The purpose of the study was the evaluation of the potential
of the algorithm for solving such a problem rather than the extraction of
 particular physics results. 

In this study a supervised statistical learning approach was followed. The algorithm was used 
to extract selection criteria for the signal/background classification
from training data samples for which it was known to which class the event belonged.
The generalisation power of the selection criteria were tested on independent data samples. 
Monte Carlo simulated data from BaBar experiment \cite{babar} was used in this study.

The input information of the algorithm was a set of event variables commonly used
in a standard cut-based analysis for the process studied:
$doca$ ( distance of closest approach between the two $\pi$ daughters of $K_S$),
$Rxy$ and $|Rz|$ (dimensions of the cylinder which defines the $e^+e^-$ interaction region) 
$|cos(\theta_{hel})|$ (absolute value of the cosine of the $K_S$ helicity angle),
 $SFL$ ($K_S$ signed flight length), $Fsig$ (statistical significance of the $K_S$ 
flight length),
$Pchi$ ($\chi^2$ probability of $K_S$ vertex) and  $Mass$ ($K_S$ reconstructed mass).
These variables, together with common mathematical functions and with floating point constants 
created by the algorithm  were used to construct the GEP chromosomes. 

The fitness function was the number of events correctly classified as signal or background.
The GEP performance in solving the problem was evaluated in terms of
classification accuracy defined as the ratio of the number of events correctly classified as 
signal or background to the total number of events in the data sample.

The GEP algorithm, using as input information only a list of logical functions and 
the event variables, developed selection criteria similar to those used in a
standard cut-based analysis, proving the algorithm works correctly:

\begin{equation}
Fsig>4.10, Rxy < 0.20cm, Pchi>0, SFL>0.20cm, doca>0cm, Rxy\leq Mass.
\end{equation} 

These selection criteria provided a classification accuracy of over 95\% for both training
and test data samples. 
The last two listed  selection criteria do not have influence on the quality of the event selection
as the inequalities are always satisfied. They are redundant information developed early
in the evolution process.

The analysis was repeated using as input functions a combination of logical and common
mathematical functions. Similar classification accuracy was obtained.
Also, no significant change of the classification accuracy was obtained by increasing 
the number of events in the data sample or by applying a parsimony pressure to the fitness
function. It was also observed that the GEP method does not suffer from overtraining
on larger data samples or on those with a larger number of event variables. These
initial results suggest GEP as a potential powerful alternative method for events
selection.

\section{Conclusions}
\hspace{1cm}Evolutionary Algorithms are, in principle, quite easy to understand computer 
algorithms. They attempt to simulate the natural evolution of species providing,
however, a quite simplistic version of this very complex  process.
These algorithms were sucessfully applied for solving complex real-world
problems in  various fields of science and engineering. Their main advantage
is  the parallel investigation of the solution landscape and the ability
to avoid trapping in local minima. As they are not  based on a rigorous
theory, there is no way, however, to demonstrate the solutions found are
indeed the best solutions. Their solutions are good as long as they are
optimal for the problem studied.

Only a limited investigation of these algorithms has been performed
in high energy physics so far. The applications discussed in this lecture are almost 
everything existing in the literature (with the exception of GA for which not all 
available studies  were included). The results of these
applications are positive and justify further exploration. The main drawback
of EA, the high computational needs, is expected to be alleviated by the increased
computer power available these days, as well as by the development of new,
more effcients versions of these algorithms.

\end{document}